\pdfoutput=1
\documentclass[sigconf]{acmart}
\AtBeginDocument{%
  }

\setcopyright{acmcopyright}
\copyrightyear{2018}
\acmYear{2018}
\acmDOI{XXXXXXX.XXXXXXX}

\usepackage{multirow}
\usepackage{threeparttable}
\usepackage{tcolorbox}
\usepackage{afterpage}
\usepackage{wrapfig}

\begin{document}

\title{Know Where to Go$:$ Make LLM a Relevant, Responsible, and Trustworthy Searcher}


\newcommand{\aff}[1]{\texorpdfstring{$^{#1}$}{}}
\author{Xiang Shi\aff{1}, Jiawei Liu\aff{1}, Yinpeng Liu\aff{1}, Qikai Cheng\aff{1}, Wei Lu\aff{1}$^*$}

\affiliation{%
  $^1$ Wuhan University, China ({\{coding,laujames2017,yinpengliu,chengqikai,weilu\}@whu.edu.cn})\\
    \country{}
}

\renewcommand{\shortauthors}{Xiang Shi, et al.}

\settopmatter{printacmref=false, printfolios=true}
\renewcommand\footnotetextcopyrightpermission[1]{}



\begin{abstract}
  
The advent of Large Language Models (LLMs) has shown the potential to improve relevance and provide direct answers in web searches. However, challenges arise in validating the reliability of generated results and the credibility of contributing sources, due to the limitations of traditional information retrieval algorithms and the LLM hallucination problem. Aiming to create a "PageRank" for the LLM era, we strive to transform LLM into a relevant, responsible, and trustworthy searcher. We propose a novel generative retrieval framework leveraging the knowledge of LLMs to foster a direct link between queries and online sources. This framework consists of three core modules: Generator, Validator, and Optimizer, each focusing on generating trustworthy online sources, verifying source reliability, and refining unreliable sources, respectively. Extensive experiments and evaluations highlight our method's superior relevance, responsibility, and trustfulness against various SOTA methods. 

\end{abstract}





\maketitle

\section{Introduction}
Since the launch of ChatGPT, large language models (LLMs) have rapidly flourished, sparking a revolutionary wave across diverse domains such as question answering \cite{robinson2022leveraging,liu2023evaluating}, machine translation \cite{mu2023augmenting}, and AI-assisted writing \cite{yuan2022wordcraft}. The field of information retrieval has been no exception to this profound transformation. Empowered by LLMs' remarkable comprehension and generation capabilities, the research paradigm in information retrieval has swiftly shifted from a ranking-centric to a generation-centric approach \citep{sun2023learning,chen2023continual}.

In this evolving domain, the advent of sophisticated generative retrieval systems such as New Bing, WebGPT \cite{nakano2021webgpt}, and WebGLM \cite{liu2023webglm} marks a notable advancement. Researchers have traversed various pathways to leverage the robust capabilities of LLMs. Some researchers propose to improve the performance of query expansion and result ranking through LLMs' linguistic abilities \cite{mao2023large, sun2023chatgpt}, while others propose the conceptualization of LLMs as intelligent agents to emulate human browsing behaviors for complex retrieval tasks \cite{nakano2021webgpt,qin2023webcpm}. Furthermore, a faction delves into LLMs' analytical and summarization prowess, scrutinizing methodologies to craft answers based on retrieved sources \cite{feng2023knowledge}.


\begin{figure}[h]
  \centering
  \includegraphics[width=\linewidth]{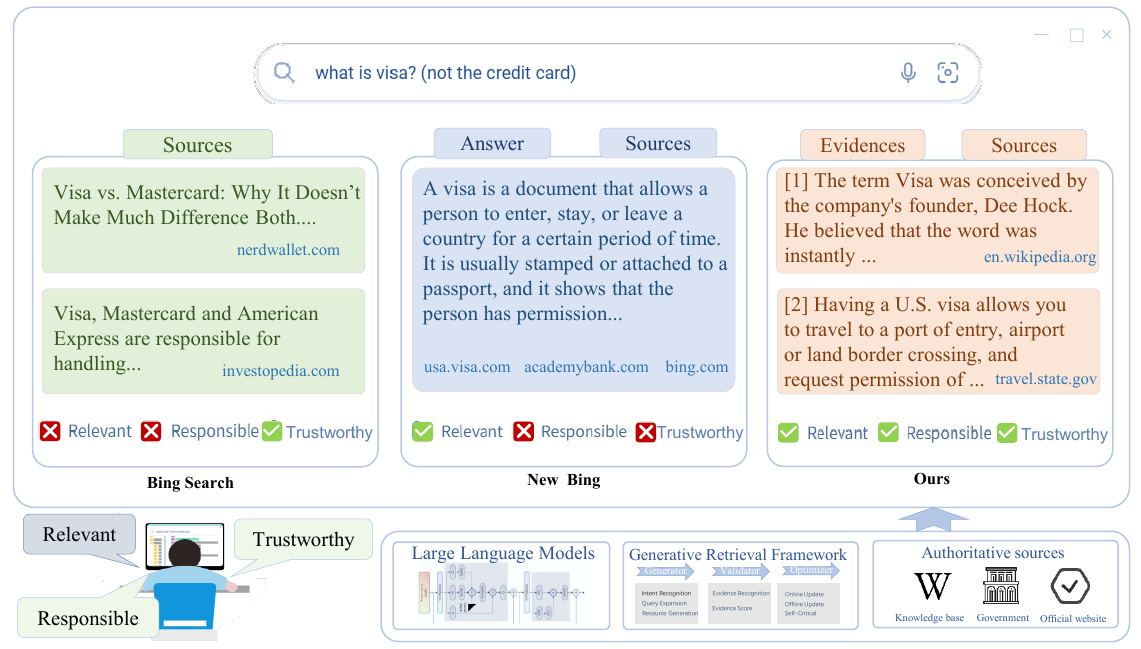}
  \caption{The performance of traditional search engines, generative retrieval approaches, and our method in responding to queries. \textnormal{Our method is capable of identifying texts that respond to queries (Relevant) from more authoritative sources (Trustworthy), and extracting reliable evidences (Responsible) from them.}}
  \label{fig:example}
\end{figure}

Despite these efforts to improve the relevance of documents returned by search engines and make LLM outputs more aligned with human preferences, a prominent challenge still remains: the reliability of sources used by generative retrieval systems cannot be guaranteed. The unreliability of these systems stems from the intrinsic limitations of both search engines and LLMs. As depicted in Figure \ref{fig:example}, search engines struggle with relevant source retrieval for vague and complex queries, a problem exemplified in the MS MARCO dataset \cite{nguyen2016ms}, where about 37.76\% of queries on Bing fail to receive effective responses. Furthermore, LLMs are constrained by inherent hallucination problems and lack accountability for the results they generate, leading to discrepancies between the generated answers and the sources they cite.

To address aforementioned challenges, we propose to construct a 'PageRank' for the LLM era, aiming to enhance the reliability of retrieval results while maintaining their relevance. Specifically, inspired by the fact that the LLMs have already acquired a wealth of web knowledge during the pre-training phase, we design a novel generative retrieval framework to harmonize the capabilities between search engines and LLMs. This framework consists three integral modules, including:


\textbf{An Intent-aware Generator:} This module primarily addresses the issue of relevance and trustfulness by leveraging the web knowledge acquired by LLMs to understand the diverse needs of users, establishing a connection between queries and online sources, and facilitating the generation of relevant and trustworthy sources.

\textbf{An Evidence-sensitive Validator}: The aim of this module is to enable the model to take responsibility for the sources it generates, by analyzing web data to verify the reliability of the sources, and to extract evidence from them that can answer the queries.

\textbf{A Multi-strategy Supported Optimizer:} This module aims to further ensure the reliability and trustfulness of the generated sources, by combining LLM's self-critique ability with web analysis capability, achieving automatic updates of invalid sources.

Extensive experiments demonstrate that our approach is capable of achieving more reliable and credible source localization and retrieval on a relative smaller-scale model (7B). Compared to existing advanced generative retrieval methods, our method exhibits a 2.54\% improvement in the validity of the recalled sources and a 1.05\% enhancement in the precision of the identified evidence.

In particular, we make the following contributions:

\begin{itemize}
\item We propose a new generative retrieval framework, designed with a generator, validator, and optimizer to respectively ensure the relevance, responsibility, and trustfulness for each retrieved source. Supported by this framework, users are empowered to achieve their retrieval objectives and acquire answers with both expediency and efficiency.

\item We introduce a multi-strategy fusion mechanism, encompassing generation, verification, and optimization, to enhance the reliability of the retrieval results. This mechanism adeptly integrates the model's self-critical capabilities with the robustness of web analysis.

\item We propose a comprehensive evaluation framework to validate the efficacy of our method. Extensive experiments demonstrate that our approach, even with a smaller parameterized model, exhibits a more discerning understanding of where to locate sources that can fulfill user requirements.
\end{itemize}

\section{Related Work}
Designing a trustworthy generative retrieval method necessitates confronting challenges like enhancing LLMs' retrieval capabilities for source evaluation, extracting query-responsive content from extensive web information, and maintaining method stability despite web page alterations. In this section, we succinctly discuss the current state of LLM development and the solutions existing research offers for these challenges.

\subsection{Large Language Models}

\textbf{(1) General models.} Since the inception of ChatGPT, a myriad of LLMs have been open-sourced by both academic and industrial communities to further research in this field. Notably, models such as LLaMA \cite{touvron2023llama}, Falcon \cite{penedo2023refinedweb}, Bloom \cite{scao2022bloom}, and others, with parameters varying from 7B to 176B, have been made available. 

\noindent\textbf{(2) Domain-specific models.} Leveraging these open-source LLM foundations, researchers have engineered specialized models aimed at diverse applications. These include Alpaca \cite{taori2023stanford} for text-based question answering, LLaMA-Adapter \cite{zhang2023llama} for multimodal question answering scenarios, Vicuna \cite{zheng2023judging} and Baize \cite{xu2023baize} for natural language chat, and Toolformer \cite{schick2023toolformer} and Gorilla \cite{patil2023gorilla} for tool invocation tasks. The performance of these models either approaches or surpasses that of ChatGPT, powered by GPT-4, within their respective application domains.

\noindent\textbf{(3) Retrieval-augmented models.} In the domain of information retrieval, researchers categorize LLMs into distinct roles such as query rewriters, retrieval enhancers, document re-rankers, and result generators. For instance, Query2Doc \cite{wang2023query2doc} amplifies the context of a query by directing LLMs to fabricate pseudo-documents, consequently elevating the likelihood of retrieving pertinent texts. LLM-URL \cite{ziems2023large} exploits the ICL capabilities of LLMs to pinpoint the URLs of corresponding pages within Wikipedia accurately. PRP \cite{qin2023large} utilizes the analytical and scoring prowess of LLMs to conduct pairwise comparisons between documents and queries, facilitating the discovery of more relevant content. ALCE \cite{gao2023enabling} evaluates the outputs generated by LLMs based on criteria such as fluency, correctness, and citation quality, while FLARE \cite{jiang2023active} implements proactive prediction coupled with multiple search engine queries to produce high-quality, long-text responses. Inspired by these studies, we devise a pipeline approach to progressively enhance LLM's retrieval capabilities, enabling direct generation and automatic validation of query-relevant sources.

\subsection{Generative Information Retrieval Systems}

Recently, researchers have developed various generative retrieval systems for scholarly purposes. Some systems focus on improving the retrieval capabilities of LLMs through the application of imitation learning. For example, WebGPT \cite{nakano2021webgpt} learns from user activities such as searches, clicks, and scrolls, thereby facilitating automated web content analysis and answer generation. Other systems, such as WebGLM \cite{liu2023webglm}, pursuit emphasize the generative capacities of LLMs, securing the accuracy and traceability of the content generated by introducing the scoring mechanism.

Apart from academic explorations, a range of commercial generative intelligent search engines like New Bing, perplexity.ai, NeevaAI, and YouChat have been launched. These engines have demonstrated significant enhancements in user experience - particularly in terms of fluidity and convenience - when contrasted with conventional search engines. However, real-world evaluations and research have identified that the verifiability and trustworthiness of responses from generative search engines still pose challenges \cite{liu2023evaluating, zhao2023can}. To address this issue, we take a distinct path from existing generative retrieval methods by refashioning LLM's retrieval approach based on source accessibility and validity, enhancing the credibility of the retrieved sources.

\begin{figure*}[h]
  \centering
  \includegraphics[width=\linewidth]{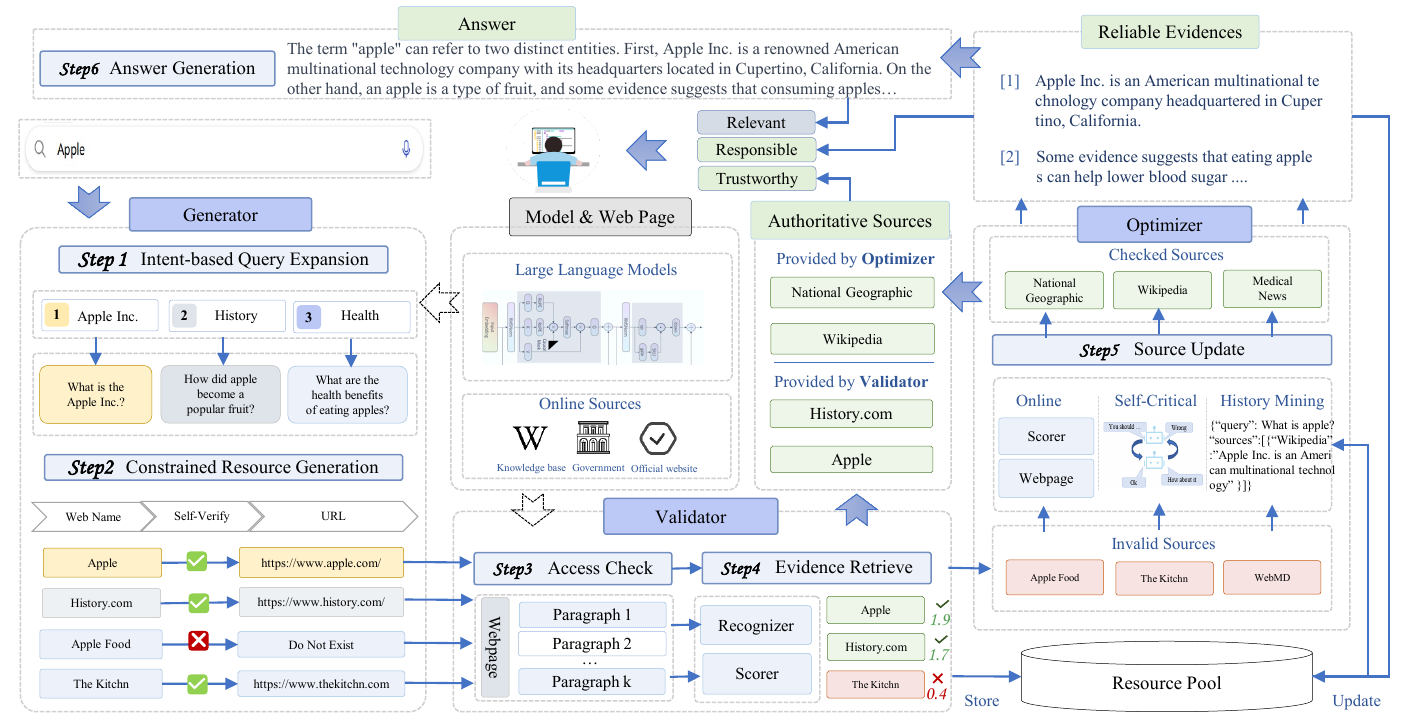}
  \caption{The overall framework of our method. \textnormal{The bottom left displays the Generator, linking query to potential sources through intent understanding and expansion. The bottom center illustrates the Validator, cross-verifying sources with real-time web data to filter credible ones. The right side depicts the Optimizer, refining unreliable sources by integrating LLM's web analysis and self-critique capabilities.}}
  \label{fig:system}
\end{figure*}

\subsection{Human Feedback and AI Feedback}

In a bid to equip LLMs with a human-like understanding of instructions and the capacity to generate non-toxic, safe, hallucination-free, informative, and creative outputs, researchers have devised the RLHF (Reinforcement Learning with Human Feedback) technique. This technique, inspired by InstructGPT \cite{ouyang2022training}, leverages a substantial volume of high-quality human preference data to direct the model's generation process. Nevertheless, the implementation of this technique incurs a substantial alignment tax.

In light of this challenge, researchers have embarked on two distinct paths. The first approach seeks to amplify the efficiency of human feedback, with strategies such as Fine-Grained HF \cite{wu2023fine}, which incorporates multiple feedback types and evaluates feedback at the sentence level, while PRM \cite{lightman2023let} emphasizes process feedback to boost the model's intermediate reasoning and generation accuracy. The second approach strives to curtail the annotation demands for human feedback data. For instance, AlpacaFarm \cite{dubois2023alpacafarm} exploits instruction data to learn and emulate human preferences, whereas RL-CAI \cite{bai2022constitutional} and PD-SA \cite{sun2023principle} employ minimal supervised signals such as self-critiquing and principles to guide the model toward automatic feedback and self-alignment. We propose a set of strategies to boost the model's self-verification and introspection abilities, ensuring stable performance amidst environmental changes by incorporating external source information.

\section{Methodology}

Existing LLM-based web search and QA systems primarily consist of three modules: the Retriever, Generator, and Scorer. In this framework, the search engine's retrieval and the model's generation are divided into retriever and generator modules, respectively. The final results are limited by the inherent constraints of both, hindering the reliability of the retrieval sources and results.

In our work, we have revamped the retrieval framework by leveraging the deep understanding of the web provided by LLMs, enabling direct association between queries and online sources. Our framework mainly comprises three modules: the Generator, Validator, and Optimizer, supported by a Source Pool retaining the query-source-evidence mapping data, as illustrated in Figure \ref{fig:system}. In the following sections, we will delve into the architecture, functionalities, and operational procedures of each module in detail.

\begin{table*}
  \caption{Comparison of URL accessibility generated by the Alpaca before and after injecting constraints. \textnormal{Query: Whiskers on cats}}
  \label{tab:constraint}
  \begin{tabular}{c||l|c||l|c}
    \toprule
    Website & URL & Exists & URL (w constraint) & Exists \\
    \midrule
    Wikipedia & https://en.wikipedia.org/wiki/Whiskers\_on\_cats & $\times$ & https://en.wikipedia.org/ & \checkmark\\ 
    WebMD & https://www.webmd.com/cats/whiskers-on-cats\#1 & $\times$ & https://www.webmd.com/& \checkmark\\
    Animal Planet & https://www.animalplanet.com/pet-care/whiskers-on-cats/ & $\times$ & https://www.animalplanet.com/& $\times$ \\
    Catster & https://www.catster.com/lifestyle/whiskers-on-cats & $\times$ & https://www.catster.com/& \checkmark\\
    PetMD & https://www.petmd.com/cat/whiskers-on-cats & $\times$ & https://www.petmd.com/& \checkmark \\
    
  \bottomrule
\end{tabular}
\end{table*}

\subsection{Generator}

Distinct from existing generative retrieval frameworks where the generator solely plays the role of answer generation, the objective of our generator is to guide the LLM in producing reliable sources, establishing a direct association between sources and queries. To achieve this aim, we set two sub-modules within the generator: one being intent-based query expansion, and the other being constrained online source generation.

\subsubsection{Intent-based Query Expansion}

In the absence of real-time web page information support, achieving direct localization of online sources necessitates fully leveraging the LLM's understanding of the interrelation between queries and sources, engaging in a step-by-step contemplation to associate queries with potential sources. To realize this, we delve into the intent underlying the queries, conducting semantic expansion thereof. As illustrated in Figure \ref{fig:system}, a query such as "Apple" embodies diverse intents - the user might be seeking information regarding Apple Inc., Apple smartphones, or the fruit. With varying interpretations of intent, the model's line of thinking gradually diverges, thereby eliciting different results.

Specifically, to thoroughly excavate the diverse user needs underlying the queries, we devise a multi-level topic generation strategy to construct intent and query expansion directive data. Our process begins with 10 broad thematic categories, which we then expand into 100 sub-themes through guiding instructions. We formulate intent recognition and query expansion instructions within each theme, ensuring the model produces evenly distributed thematic instruction data. This approach bolsters the LLM's responsiveness to a wide array of question types. The topics utilized during data generation are presented in Appendix B.

\subsubsection{Constrained Online Source Generation}

Upon enriching the semantic information of queries, it becomes imperative to gradually guide the LLM in leveraging this diversified information to unearth potentially relevant sources. The foundation for actualizing this process lies in the fact that during the pre-training phase of LLMs such as LLaMA \cite{touvron2023llama} and Falcon \cite{penedo2023refinedweb}, web data like C4 and CommonCrawl constitute 80\% and 100\% of their pretraining corpus respectively. These data retain valuable information like webpage URLs and names. Simultaneously, within the fine-tuning data of models like Alpaca \cite{taori2023stanford}, we identified a subset of data dedicated to web content recognition and link analysis, which inspired us to devise the following constrained online source generation method.

\textbf{Online Source Generation.} A two-phase instruction is employed to accurately guide the model in identifying online sources related to a given query, as demonstrated in the first row of Table \ref{tab:self_verify}. The fundamental premise of this methodology conceptualizes the LLM itself as a vast representation of the internet. The initial phase encompasses query comprehension, which mandates the model to recommend, in sequence, potential webpage names resonating with user intent, derived from an explicitly defined query. The subsequent phase involves the identification of webpage URLs, tasking the model with discerning the primary URL corresponding to each webpage name, thus facilitating direct location and access. Notably, although it is plausible to instruct the LLM to generate URLs based directly on the query, such a procedure impedes the model's capacity for understanding the nexus between queries and webpages. Moreover, this direct approach limits opportunities for introducing constraints during the generation process, minimizing the illusion effect, which will be elucidated in subsequent sections.

\textbf{Generation Constraint.} In our analysis of models such as ChatGPT and Alpaca, we note that, without any constraints, these models tend to generate extended URLs which embed the initial query phrase. While these URLs may seem logically structured, they are actually inaccessible—a phenomenon commonly referred to as the illusion problem, as illustrated in the left column of Table \ref{tab:constraint}.

While the model demonstrates the aforementioned challenges, potential utility can still be derived. By extracting solely the domain name from these URLs, we can effectively point to web sources that are highly pertinent to the original query. As illustrated in the right column of Table \ref{tab:constraint}, a random sampling of 100 queries from the MS MARCO dataset confirms that in an unconstrained setting, the actual accessibility of URLs produced by the Alpaca model stands at a mere 39.25\%. Yet, with the imposition of appropriate constraints, this accessibility surges to a significant 71.07\%.


\begin{table}
  \caption{The example of constrained online source generation process. \textnormal{Integrating self-verification with constraints to alleviate the illusions of LLM.}}
  \label{tab:self_verify}
  \begin{tabular}{l|p{6cm}}
    \toprule
    Step & Example \\
    \midrule
    Query $\rightarrow$ Web & What are the health benefits of eating apples? $\rightarrow$ 1.Healthline 2. Apple Food \\
    \midrule
    \multirow{2}{*}{Web $\rightarrow$ URL} & Healthline $\rightarrow$ https://www.healthline.com/ \checkmark\\ 
     & Apple Food $\rightarrow$ Do Not Exist $\times$\\
    
  \bottomrule
\end{tabular}
\end{table}

\textbf{Self-Verification.} Owing to challenges such as illusions and outdated data, the model's outputs can be categorized into three distinct types: 1) The generated website name does not exist (indicating an generative illusion); 2) The generated website name exists, but the corresponding URL is either non-existent or does not match the website name (suggesting an illusion or URL is changed); 3) The generated website name and URL are accurate, but the webpage does not contain content relevant to the user's query (another form of illusion). To tackle the first two issues, we introduce a self-verification strategy, as shown in the last row of Table \ref{tab:self_verify}. We instruct the model to perform a secondary existence check on the website when generating URLs based on the website name. As for validation and optimization methods concerning the third issue, please refer to sections 3.2 and 3.3.

\subsection{Validator}

The Validator takes the sources generated by the Generator as input, invoking the LLM's webpage analysis capability to examine the timeliness, accessibility, consistency, and validity of the generation results. Centered around these indicators, we design an automated verification method to incrementally access webpage information, directly extracting evidence sentences from the webpages that can answer the queries as outlined within the sources.

The automatic evaluation procedure can be expressed by the following equation. Assuming there are $n$ implicit intentions $I = \{I_1, I_2, ..., I_n\}$ in user query $Q$, and based on each intention, the generated expansion queries $EQ$ can individually retrieve $r$ online sources with URLs. The first step is to feed the $R$ URLs, as generated by the LLM, into the search engine to validate the timeliness and accessibility of these sources.
\begin{equation}
   R = \bigcup\limits_{i=1}^n\{R_{i1}, R_{i2}, ..., R_{ir}\}
\end{equation}

\begin{equation}
    R^* = AccessCheck(R)
\end{equation}

After eliminating nonexistent generated sources, we obtain a refined list comprising $R^*$ accessible sources. We then concatenate the model's $n$ expanded queries with the $|R^*|$ sources to formulate query expressions for advanced web page retrieval.

\begin{equation} 
    EQ^* = \bigcup\limits_{i=1}^n{Concate[EQ_i:R^*_i]}
\end{equation}

The refined query expressions are subsequently entered into Bing's search interface to retrieve the top $K$ search outcomes. Through web parsing utilities, we determine the accessibility of these web pages and extract textual data from the accessible ones.

\begin{equation}
    T = API_{top_k}(EQ^*)
\end{equation}

After extracting the web page text $T$, we segment the content using fixed windows (length=$m$) $T=\{T_{1}, T_{2},...,T_{|T|/m}\}$. Subsequently, we employ an LLM-based evidence retriever to extract the evidence sentences $T^*$ from the text $T$ that can address the query.

\begin{equation}
    T^*=Evidence(T)
\end{equation}

Inspired by \citep{sachan2022improving}, we introduce two evidence retrieval modules into our evidence retriever. Firstly, an evidence recognition module, which identifies fine-grained evidence sentences capable of answering queries from extensive text through a generative manner. Secondly, an evidence scoring module, which evaluates the reliability of each text segment through scoring, considering them as valid evidence if the score surpasses a threshold. The score for the evidence stems from the LLM's probability assessment regarding the text's ability to address the queries, characterized as 'Yes/No'.


\begin{table}
  \caption{Data augmentation. \textnormal{We augment the dataset by modifying queries and corresponding evidence sentences to introduce strong negative cases and diversify positive cases.}}
  \label{tab:data_agument}
  \resizebox{0.48\textwidth}{!}{
  \begin{tabular}{l|p{6cm}}
    \toprule
    Operation & Description\\
    \midrule
    $Shuffle$ & Alter the position of the evidence sentences\\ 
    $Expand/Simplify$ & Add or remove irrelevant sentences\\
    $Rephrase_q$ & Modify the query to a related query\\
    $Rephrase_{q^-}$ & Modify the query to an unrelated query\\
    $Rephrase_d$ & Alter the expression of the evidence sentences \\
    $Complexify$ & Transform the query into a complex query\\
  \bottomrule
\end{tabular}
}
\end{table}

\begin{equation}
    Score(T_j) = \begin{cases}
    1+P(Yes|T_j), & T_j \in T \ \& \ Output=Yes \\
    1-P(No|T_j), & T_j \in T \ \& \ Output=No
    \end{cases}
\end{equation}

Based on these two modules, we design two evidence retrieval strategies. The first is a score-only strategy, utilizing solely the scoring module to record evidence sentences as they are identified. The advantage of this strategy is its ability to recall more evidence sentences, albeit with a coarser granularity. The second is a hybrid strategy, initially employing the recognition module to fine-tune the extensive text, followed by a secondary verification of the identified evidence using the scoring module. This strategy yields more precise identification results, though illusions may arise during model-generated evidence. To mitigate such occurrences, we adopt a series of data augmentation schemes as depicted in Table \ref{tab:data_agument}.

\subsection{Optimizer}

The objective of the source Optimizer is to dynamically adapt the model's source generation in response to evaluation outcomes, thereby aligning with user requirements. This is achieved through our multi-strategy source optimization method, consisting of self-critical, online, and history mining approaches.


\textbf{Self-Critical Strategy} We leverage the model's self-verification capability for web source generation, automatically updating the source list based on source accessibility and relevance. Taking the generation results of the model in Table \ref{tab:self_verify} as an example, after verification by the model itself, it was found that the "Apple Food" webpage does not exist. Therefore, we convert this verification result into a natural language instruction input to the model, asking the model to replace the non-compliant websites in order. After self-verification, the model outputs the results as "1. Healthline 2. WebMD". Once the model completes the updates, the newly generated websites are sent back to the validator for further verification.

\textbf{Online Strategy} To integrate the strengths of search engines with the features of LLM, we utilize the previously designed evidence verification module. This aids in exploring and discovering webpage sources, maximizing the retrieval of web content closely associated with user queries.

\textbf{History Mining Strategy} Inspired by the studies in recommendation system, another efficient approach is to mine potential sources from historical records. Online source mining based on historical information are conducted by the following three steps.

(1) Pre-population of the Source Pool: The source pool serves as the foundation for implementing history mining strategies, housing the mapping data of query-source-evidence. When the source pool lacks sufficient data, pre-population of the resource pool is necessitated. To accomplish this task, we employ the data augmentation schemes mentioned in Table \ref{tab:data_agument} to transform queries, and secure a certain number of retrieval results through the validation module.

(2) Similarity Calculation: As our source pool accumulates a significant volume of data, the retrieval of new sources becomes a matter of determining similarity. We gauge the proximity between the user's query and existing queries within our source pool. When this similarity score exceeds a preset threshold $\delta$, corresponding sources are populated to candidate list $R_{cand}$ for further validation.

\begin{equation}
    R_{cand} = \{Similar(Q, Q^*)>\delta| \forall Q^* \in Source Pool\}
\end{equation}

(3) Re-verification: The candidate sources, along with the query, are re-submitted to the validator. Valid sources are updated to the retrieval results. This process continues until the candidate list is exhausted, or the desired source limit is attained.

    

\section{Experiment}
In this section, we conduct a detailed comparison between the performance of our method and existing generative search engines and technologies. Subsequently, a comprehensive evaluation is carried out on various modules within our retrieval framework such as the generator, validator, and optimizer, as well as specific tasks. Moreover, we analyze how our method operates in conjunction with search engines, achieving improvements on relevance, responsibility, and trustworthiness.

\subsection{Evaluation Criteria}

\begin{table*}[htbp]
  \caption{Performance Evaluation. \textnormal{The left-side metrics display the count and topic characteristics. The right-side metrics reflect the reliability of the retrieval results. Values in brackets indicate performance after removing duplicate sources from retrieved results.}}
  \label{tab:score}
  \begin{tabular}{ccccc|ccccc}
    \toprule
    \multirow{2}{*}{System} & \multicolumn{4}{c}{Statistical Metrics} & \multicolumn{5}{c}{Performance Metrics} \\
\cmidrule{2-10}
     & $S_{count}$ & $E_{count}$ & $Q_{correct}$ & $T_{avg}$ & Timeliness $\uparrow$ &Access $\uparrow$ &Consistency $\uparrow$& Validity $\uparrow$ &Precision $\uparrow$\\
    \midrule
    New Bing & 903 (460) &  427 & 28 &1.59 & 97.56 (99.57) & 97.56 (99.57) & 97.56 (99.57) & 73.53 & 72.83 \\
    Perplexity.ai & \textbf{1595 (1038)} & 1107 & 42 &\textbf{2.42} & 99.37 (99.33) & 99.24 (99.23) & 99.31 (99.23) & 73.35 & 67.57 \\
    WebGPT (175B) & 950 (505) & 950 & 154 &1.68 & 97.15 (96.83) & 96.94 (96.44) & 96.73 (96.23) & 84.63 & 77.36 \\
    WebGLM (10B) & 1355 (513) & \textbf{1355} & 152 &1.76 & 99.55 (99.81) & 99.40 \textbf{(99.81)} & \textbf{99.40 (99.81)} & 85.38 & 74.83 \\
    \midrule
    Our Method (7B) & 295 (173) & 565 & \textbf{178$^*$} & 1.29 & \textbf{100.00}  & \textbf{99.81}(99.66)  & 97.21(95.62) & \textbf{87.92$^*$}  & \textbf{78.41$^*$} \\
  \bottomrule
\end{tabular}
\end{table*}

\subsubsection{Data}

To rigorously evaluate the performance of our method, we employ two distinct datasets. For an impartial assessment of the retrieval capabilities, we utilize a subset of the ELI5 dataset \cite{fan2019eli5}, which was curated in prior work \cite{nakano2021webgpt}. To offer a granular examination of the performance attributes across various modules and sub-tasks within the framework, we construct a comprehensive evaluation dataset that spans all functional modules and tasks. Detailed methodologies for dataset construction, as well as illustrative examples, are provided in Appendix B.

\subsubsection{Metrics}

We introduce four statistical metrics and five performance metrics to systematically quantify the performance disparities between source-generation systems and source-retrieval systems. The four statistical metrics include online sources count ($S_{count}$), evidence sentences count ($E_{count}$), queries yielding correct responses ($Q_{correct}$), and average number of topics for which results are retrieved per query ($T_{avg}$). The performance metrics are delineated as follows:

\begin{itemize}
\item \textbf{Timeliness}. Evaluate if the model-generated web pages (by name) exist online, accounting for possible renaming over time. Web page names from existing retrieval systems are represented by domain names.

\item \textbf{Access}. Evaluating the accessibility of URLs that are either retrieved or generated by the system.

\item \textbf{Consistency}. Investigating the relevance of the online sources (by name) either retrieved or generated by the system to the corresponding URLs.

\item \textbf{Validity}. Quantifying the ratio of sources that effectively address queries among the entire set of web sources either retrieved or generated by the system.

\item \textbf{Precision}. Ascertaining the sufficiency of evidence sentences automatically extracted from web sources by the system in substantiating responses to queries.
\end{itemize}

For ease of description, we adopt the first four characters of the performance metrics as abbreviations. The up arrow $\uparrow$ indicates that a higher number in the metric signifies better performance. $*$ denotes significant results.

\subsubsection{Baseline}

We utilize four popular commercial and open-source generative retrieval systems to serve as our experimental baselines. 
\begin{itemize}
\item \textbf{New Bing:} an advanced generative retrieval system based on GPT-4 and Bing Search.
\item \textbf{Perplexity.ai:} A generative retrieval system developed on the foundation of GPT-3.5.
\item \textbf{WebGPT:} a generative retrieval model based on the GPT model, trained using user search behavior data. 
\item \textbf{WebGLM:} an open-source generative retrieval system based on the GLM model, incorporating lightweight evidence identification models such as Contriver.
\end{itemize}

\subsubsection{Method}

To ensure the accuracy of the evaluation results, we recruit 10 experts with a master's degree or higher in the field of information retrieval to conduct the human evaluation.




\subsection{Comparison}

Table \ref{tab:score} presents a comprehensive comparison between our method and existing systems, employing both statistical and performance metrics for assessing retrieval capabilities.

The left four columns of Table \ref{tab:score} show that search-based systems have an advantage in online sources and evidence recall. However, more recalled sources don't guarantee increased answer accuracy. Expert evaluation shows that systems like New Bing and Perplexity.ai, despite diverse source retrieval, include accurate evidence in less than 20\% of responses. In contrast, our method and specialized systems like WebGLM and WebGPT provide more reliable results (55\% $-$ 65\%) due to their evidence identification engineering. Additionally, our training's topic-centric approach grants better topical coherence between generated sources and queries, unlike other systems mainly focusing on relevance, often matching a single query to multi-topic webpages. This topic consistency to some extent reflects the relevance of our generated results to the queries.

The rightmost five columns of Table \ref{tab:score} show that search-based systems don't achieve a perfect score (100\%) in metrics like timeliness, accessibility, and consistency, highlighting the dynamic nature of online sources from retrieval to evaluation. Our method, owing to the incorporation of the generation step, still falls short in consistency. Nevertheless, with the aid of specialized validators and optimizers, our method demonstrates significant strides in the validity and precision of recalled sources and evidences, achieving scores of 89.72\% and 78.41\%, respectively. This indicates that our method achieves higher relevance in the retrieved sources and can better account for its retrieved results.

\subsection{Ablation Study}
We conduct ablation studies at both the module and task levels. Table 5 reveals that, in the absence of validation and optimization modules, our method attains a timeliness, accessibility, and consistency of 94.24\%, 82.98\%, and 76.44\%, respectively. This suggests that over 75\% of the generated sources are immediately utilizable by search engines. Furthermore, even without appropriate validation, the relevance of these generated sources to the query stands at 58.89\%, underscoring the reliability of the model-generated online sources. Upon the incorporation of validation and optimization modules, there is a significant uptick across all metrics. Notably, our method exhibits higher validity without the optimizer, a phenomenon attributed to the inadvertent inclusion of irrelevant sources during the online retrieval and validation phase due to limitations in the validator's performance. Additionally, inspection of the first two columns of Table 6 reveals that under the regimen of multi-task joint fine-tuning, the performance of the generator remains largely invariant, while the validator experiences a performance enhancement ranging between 5\% and 10\%. 

\begin{table}

  \caption{Module-level ablation study. \textnormal{Reliability of retrieval sources without using validator and optimizer.}}
  \label{tab:topic}
  \resizebox{0.48\textwidth}{!}{
  \begin{tabular}{cccccc}
    \toprule
    & Timeliness $\uparrow$ &Access $\uparrow$ &Consistency $\uparrow$& Validity $\uparrow$ &Precision $\uparrow$\\
    \midrule
    Full &  \textbf{100.00} & \textbf{99.81(99.66)} & 97.21 \textbf{(95.62)} & 87.92 & \textbf{78.41}\\
   w/o opt. & \textbf{100.00} & 99.51 (98.77) & \textbf{98.05} (95.06) & \textbf{88.76} & 67.56 \\ 
   w/o val. & 96.94 (94.24) & 89.44 (82.98) & 85.83 (76.44) & 58.89 & $-$ \\

  \bottomrule
\end{tabular}
}
\end{table}

\subsection{Capability Evaluation}

We conduct a systematic evaluation to ascertain the variations on the source generation, validation, and optimization capabilities under diverse parameter configurations and strategic selections.

\subsubsection{Source Generation Capability.}

We design a tripartite experimental framework to rigorously evaluate the model's source generation capabilities. The first experiment is designed to assess the performance metrics within the source generation module; the second probes the sensitivity of the model's generative prowess to parameter configurations; and The third aims to examine the trustfulness of the model-generated results and their differences from search engines.

\begin{table}
  \caption{Task-level ablation study. \textnormal{Union, Sep, and Sep$_{one}$ represent the results of multi-task joint tuning, single-task tuning, and single-task tuning using multi-task parameters, respectively. Fuzzy Match refers to predictions encompassing the gold evidence, while Exact Match requires predictions to match the standard exactly.}}
  \label{tab:basic}
  \resizebox{0.4\textwidth}{!}{\begin{tabular}{ccccc}
    \toprule
   Metric & Task & Union & Sep & Sep$_{one}$\\
    \midrule
    \multirow{4}{*}{Accuracy} & Intent Recognition & 62.20 & \textbf{63.33} & 38.40  \\
     & Query Expansion & 74.30 & \textbf{76.80} & 76.80  \\
    & URL Identification & 57.00 & \textbf{57.30} &  45.80 \\
    & Web Identification & 82.51 & \textbf{82.88} &  64.30 \\
    \midrule
   \multirow{2}{*}{Exact Match} & Evidence Recognition & \textbf{51.20} & 46.32 &  28.38 \\
   & Evidence Scoring & \textbf{47.43} & 44.39 &  34.38 \\
   \midrule
   \multirow{2}{*}{Fuzzy Match} & Evidence Recognition & \textbf{65.67} & 55.22 &  35.19 \\
   & Evidence Scoring & \textbf{82.15} & 78.81 &  35.10 \\
  \bottomrule
\end{tabular}}
\end{table}

\begin{table*}
  \caption{Source generation capability evaluation. \textnormal{$n$ and $r$ respectively denote the hyperparameters used by the generator to control the diversity and quantity of source generation. Validity of generated sources remains similar with or without using the original query.}}
  \label{tab:generation}
  \resizebox{0.85\textwidth}{!}{
  \begin{tabular}{cc|cccc|cccc}
  \toprule
    \multicolumn{2}{c}{Parameters} & \multicolumn{8}{c}{Metrics}  \\
    \midrule
  & & \multicolumn{4}{c}{Use Original Query} & \multicolumn{4}{c}{Not Use Original Query} \\
\cmidrule{3-10} 
   n  & r &Timeliness $\uparrow$ &Access $\uparrow$ &Consistency $\uparrow$& Validity $\uparrow$ &Timeliness $\uparrow$ &Access $\uparrow$ &Consistency $\uparrow$& Validity $\uparrow$\\
    \midrule
   1 & 1 & 96.60 (93.47) & 88.74 (79.90) & 85.60 (74.87) & 59.69  & 97.79 \textbf{(96.05)} & 88.92 (81.88) & 85.61 (76.32) & 58.67\\
   1 & 2 & 96.64 (92.66) & 88.99 (80.70) & 84.70 (73.09) & 57.05 & 97.60 (94.84) & 90.20 (83.81) & 85.58 (75.40) & 56.38\\
   1 & 3 & \textbf{97.34 (93.68)} & 90.82 (82.21) & \textbf{86.96} (75.17) & 57.30 & \textbf{98.01} (95.36) & \textbf{92.04 (85.45)} & \textbf{88.18 (77.68)} & 57.10\\
   2 & 1 & 96.34 (93.21) & 89.24 (81.94) & 86.04 (76.47) & \textbf{59.95} & 96.94 (94.24) & 89.44 (82.98) & 85.83 (76.44) & 58.89\\
   2 & 2 & 96.13 (91.76) & 89.33 (81.23) & 84.53 (73.14) & 56.86 & 96.76 (92.81) & 90.27 (83.77) & 85.75 (75.31) & 56.98\\
   2 & 3 & 96.26 (91.44) & 90.38 (81.33) & 85.45 (73.57) & 55.64 & 96.30 (91.48) & 91.17 (83.53) & 86.04 (74.89) & 55.27\\
   3 & 1 & 96.50 (93.47) & 89.09 (82.20) & 86.01 (\textbf{76.73}) & 59.88 & 96.93 (94.04) & 89.13 (83.02) & 85.58 (76.15) & \textbf{59.10}\\
   3 & 2 & 96.19 (91.83) & \textbf{89.64 (82.40)} & 84.25 (72.84) & 57.29 & 96.59 (92.49) & 90.39 (84.46) & 84.79 (73.99) & 57.54\\
   3 & 3 & 96.43 (91.77) & 90.44 (81.41) & 84.95 (73.07) & 55.21 & 96.51 (91.86) & 91.07 (83.30) & 85.23 (73.67) & 55.52\\
  \bottomrule
\end{tabular}
}
\end{table*}

(1) Basic Capability Test. As evidenced by the first and third columns of Table \ref{tab:basic}, the model exhibits a suboptimal URL identification rate of 57\%. This shortcoming is primarily due to the inaccurate categorization of non-existent websites. Conversely, the recognition accuracy for existing websites escalates to 76.21\%. Moreover, marked performance variances emerge when subjecting individual tasks to isolated versus coordinated training, corroborating the model's capability to discern the latent interdependencies among these retrieval sub-tasks.

(2) Generation Stability.
The quantity of sources ultimately generated by the source generation module is primarily determined by the number of intent judgments $n$, and the number of web results $r$ generated for each query. We adjust these two parameters to vary between 1 and 3, recording the timeliness, accessibility, consistency and validity of the generated sources. As evidenced in Table \ref{tab:generation}, our empirical findings substantiate that the model's source-generation capabilities maintain a level of stability even when the volume and diversity of the generated sources are incrementally increased. This observation serves as a tacit validation of the model's efficacy in ameliorating the 'illusion problem,' inherently present in LLMs, during the process of online source generation.

\begin{figure}[!t]
  \centering
  \includegraphics[width=\linewidth]{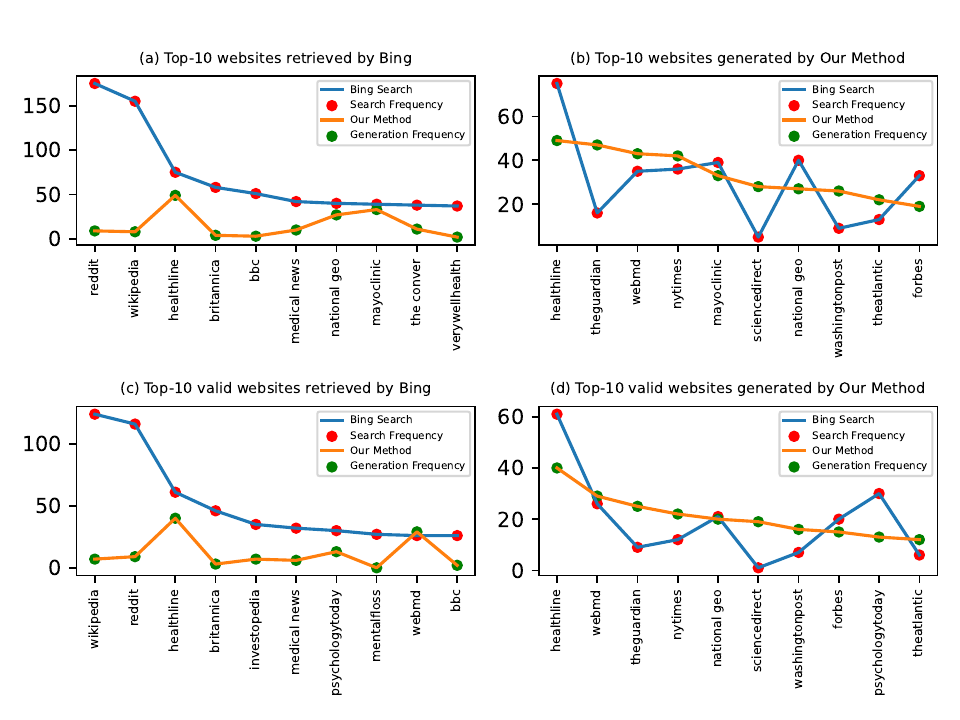}
  \caption{Top-10 frequently retrieved and generated sources by Bing and our method. \textnormal{The X-axis represents the website names, and the Y-axis represents the frequency.} }
  \label{fig:distribution}
\end{figure}

(3) Trustfulness of Generated Sources. Figure \ref{fig:distribution} displays the distribution of sources retrieved and generated by Bing Search and our model, respectively. From sub-figures (a) and (b), it is apparent that although there are some disparities in the distribution between the search engine results and model-generated results, among the top 10 generated by the model, 3 appear in the top 10 results of the search engine, and 9 appear in all the results returned by the search engine, with the most frequent, "Healthline," appearing in the 3rd position of the search results. When considering only valid sources in sub-figures (c) and (d), the distribution between the two further converges, with sources such as "WebMD" and "National Geographic" exhibiting nearly identical frequencies. Additionally, when solely considering the authority of the model-generated sources, 7 out of the top 10 results belong to authoritative media and journal publications, while 3 belong to renowned medical institutions and knowledge bases. These results suggest a consistency in trustfulness between sources generated by the model and those recommended by the search engine, indicating that the model-generated results can complement the search engine retrieval outcomes.

\begin{table}
  \caption{Source validation capability evaluation. \textnormal{${}^{*}$ indicates statistical significance at $p<$ 0.05 level compared to the best performance of baselines. Contriever$^{\dagger}$ refers to the retrained model in the work of \citet{liu2023webglm}.}}
  \label{tab:evidence}
  \resizebox{0.42\textwidth}{!}{
  \begin{tabular}{c|ccc|c}
    \toprule
    \multirow{2}{*}{Method} & \multicolumn{4}{c}{Metrics}  \\
    \cmidrule{2-5}
    & \multicolumn{3}{c}{Ranking} & Recognition\\
\cmidrule{2-5}
    &Accuracy&NDCG&NDCG@5&Accuracy\\
    \midrule
   TF-IDF & 9.70 & 87.43 & 82.12 & 47.68\\
  BM25 & 9.75 & 83.55 & 76.15 & 41.57\\
  Contriever & 12.15 & 87.06 & 87.06 & 72.90\\
   Contriever$^{\dagger}$ & 12.26 & 86.69 & 86.69 & 72.77\\
   \midrule
  Ours & \textbf{13.35}* & \textbf{92.81}* & \textbf{90.06}* & \textbf{77.05}*\\
  \bottomrule
\end{tabular}}
\end{table}

\subsubsection{Source Validation Capability.}

In testing the source validation capability, two distinct evaluations are conducted: evidence recognition and evidence ranking. The former measures the model's precision in isolating granular evidence sentences that are pertinent to queries extracted from extensive texts. The latter quantifies the model's ability to re-verify the credibility of identified evidence sentences. As illustrated in Table \ref{tab:evidence}, As delineated in Table 8, relative to Contriever, our method achieves a 4.15\% enhancement in its evidence-recognition capabilities. Also, the validator manifests incremental gains across each metric in ranking performance, specifically advancing by 1.09\%, 5.38\%, and 3\% respectively. 

\begin{table}
  \caption{Performance of optimizer using different strategies.}
  \label{tab:optimzer}
  \resizebox{0.48\textwidth}{!}{
  \begin{tabular}{cccccc}
    \toprule
    & Timeliness & Access &Consistency & Validity & Precision\\
    \midrule
   Critical & 100.00 & 99.20 (98.36) & \textbf{99.20 (98.36)} & \textbf{89.09} & 71.48 \\
   Online & 100.00 & \textbf{99.81 (99.66)} & 97.21 (95.62) & 87.92 & \textbf{78.27}  \\ 

  \bottomrule
\end{tabular}
}
\end{table}

\subsubsection{Source optimization Capability.}

To rigorously evaluate the source optimization capabilities of our method, we juxtapose its performance under self-critical and online strategies. It's worth noting that the history mining strategy's effectiveness is contingent upon the quality of the source pool, for which exemplifications are provided in Appendix C. As evidenced by Table \ref{tab:optimzer}, the self-critical strategy of the model yields reliable online sources and demonstrates superior relative validity as compared to the online strategy. However, it underperforms in terms of evidence recognition precision relative to the online approach.

\subsection{Model Scaling}
\begin{figure}[h] 
  \centering
  \includegraphics[width=\linewidth]{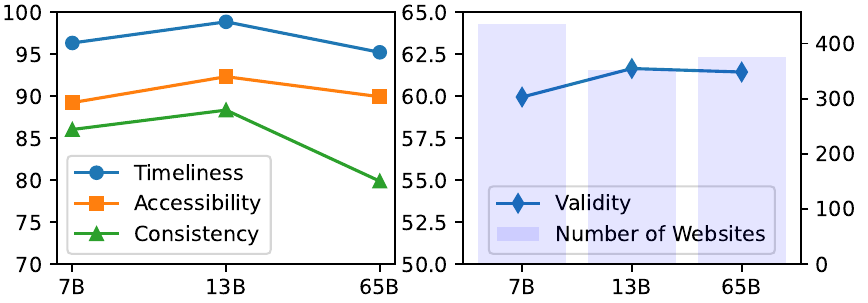}
  \caption{The variation in the number and reliability of sources generated by the model with the increase in model parameters.}
  \label{fig:scale}
\end{figure}
As illustrated in Figure \ref{fig:scale}, when the model parameters are elevated from 7B to 13B, there is a gradual enhancement in metrics such as timeliness, accessibility, consistency, and validity. Upon further escalation to a 65B parameter, the model produces a more diversified collection of sources, especially specific journal magazines like "Journal of Chemical Education" and "Frontiers in Psychology". However, these magazines do not possess independent URLs on the web but are housed under the respective publisher's websites, which leads to a decrease in the consistency of the sources generated by the model. Nonetheless, the validity of the sources produced by the 65B model remains unaffected, and compared to the 13B model, it can generate more reliable sources.

\section{Conclusion}

In this paper, we introduce a generative retrieval framework to evolve the LLM into a relevant, responsible, and trustworthy searcher. Through well-designed generation, verification, and optimization strategies, the LLM mines pre-training web knowledge effectively, even without visible web page information, linking queries directly to reliable online sources. The addition of subsequent web page information aids in evidencing each valid source, providing a foundation for the generated results. Extensive experiments show that our framework notably outperforms previous methods in source retrieval reliability, opening a new pathway for generative search engine development.


\bibliographystyle{ACM-Reference-Format}
\bibliography{literature.bib}

\appendix

\section{Training and Inference}

\subsection{Training Details}
\subsubsection{Model} 
We utilize the popular open-source LLM Llama-7B as the base model for fine-tuning the source retrieval model. In addition, we train models with 13B and 65B parameters and conduct an extensive performance analysis across these varying model complexities.

\subsubsection{Hyperparameters}
The model training is conducted with an epoch setting of 1, a batch size of 64, an initial learning rate of 3e-4, a weight decay of 0.1, and adapters are added after all attention weights and MLP layers in the base model.

\subsection{Inference Details}
\subsubsection{Parameter Configuration}
The model's inference is chiefly guided by two parameters: the number of intents ($n$) and the number of web pages ($r$). Typically, $n=2$ and $r=1$ are set, yielding satisfactory source generation efficiency. However, for source distribution analysis, settings are adjusted to $n=3$ and $r=3$, aligning generated sources with the first page of search engine results, enabling a robust comparison. The maximum allowable values for $n$ and $r$ are 5. The thresholds for evidence scoring and historical mining strategy are 1.7 and 0.5, respectively.

\subsubsection{Training and Inference Prompt}
The designated prompt for inference is structured as follows: herein, {Module} demarcates the involved model component (be it generator, verifier, or optimizer), {Task} elucidates the precise instruction to be executed within the module (such as intent recognition, query expansion, source retrieval, etc.), {Input} denotes the model's input, and {Output} signifies the model's output.

\begin{tcolorbox}[colback=gray!10, boxrule=0pt]
You are Know Where to Go, an intelligent source retrieval system composed of a generator, validator, and optimizer. In the generator, your tasks are as follows: 1. Intent Recognition. 2. Query Expansion. 3. Source Retrieval. 4. URL Identification. 5. Website Identification. In the validator, your tasks are as follows: 1. Evidence Recognition. 2. Evidence Check. 3. Answer Generation. In the optimizer, your primary task is as follows: 1. Source Correction.
\\ \\
Below are the system modules and task descriptions. Please respond accurately to the system functionalities.
\\ \\
\textbf{\#\#\# Module:}
\{Module\}

\textbf{\#\#\# Task:}
\{Task\}

\textbf{\#\#\# Input:}
\{Input\}

\textbf{\#\#\# Output:}
\{Output\}
\end{tcolorbox}

\section{Dataset Example}


In our study, we assemble a dataset of 54,200 instructions categorized into: intent recognition (3.8k), query expansion (4.8k), source generation (11k), URL identification (3.4k), Web name identification (3.8k), evidence recognition (4.5k), evidence check (5.4k), source correction (2.5k), and answer generation (15k). While answer generation isn't our focus, we adapt a dataset from the WebGLM study for it. For other categories, we merge GPT-3.5-turbo and MSMARCO dataset with manual curation, employing a hierarchical method for collecting intent-based instructions. This forms the base for formulating query expansion and source generation instructions, ensuring thematic diversity, elaborated in Table \ref{tab:theme}. We specifically augment data for evidence recognition instruction set to enrich diversity, detailed in Table \ref{tab:data_agument}. Examples of instructions for these nine tasks are showcased below.

\begin{table*}
  \caption{Examples of Themes used to generate intent recognition instructions.}
  \label{tab:theme}
  \begin{tabular}{c|c}
    \toprule
   Main Topic  & Sub Topics\\
    \midrule
   Health and Lifestyle & Exercise, Nutrition, Mental Health, Sleep, Stress Management, Disease Prevention \\
   Finance and Investment & Personal Finance, Tax Planning, Credit Cards, Loans, Stocks \\
   Business and Marketing & Business, Economy, \\
   Computers Science and Technology & Artificial Intelligence, Machine Learning, Robotics, Virtual Reality, Augmented Reality\\
   Creativity and Design & Kitchen Design, Bathroom Design, Color Schemes, Gardening, DIY Home Improvement\\ 
   Entertainment and Culture & Movies, Plays, Theater, Pop Culture, Festivals\\ 
   Travel and Vacation &Adventure Travel, Backpacking, Beach Holidays,  City Breaks, Cruise Travel\\
   Sports and Fitness & Running, Walking, Swimming, Cycling, Hiking\\
   Technology and Electronics & Computer Hardware, Wi-Fi, 5G Technology, Networking, Digital Assistants\\
   Food and Beverages & Cooking, Baking, Meal Planning, Food Safety, Coffee \\
  \bottomrule
\end{tabular}
\end{table*}

\subsection{Generation Instruction}
\subsubsection{Intent Recognition Instruction}
The objective of intent recognition instruction is to identify the user's underlying $n$ intents (where $1 \leq n \leq 5$) within a query. The input is the query sentence, and the output consists of a specified number of intents.

\begin{tcolorbox}[colback=gray!10, boxrule=0pt]
\textbf{Module:} generator

\textbf{Task:} [Intent Recognition] Recognizing two intents based on user's query.

\textbf{Input:} [Query] what is production

\textbf{Output:} [Intent] 1. Definition of production 2. Types of production
\end{tcolorbox}

\subsubsection{Query Expansion Instruction}
The goal of the query expansion instruction is to externalize the latent intent of a query, transforming the original query to better retrieve relevant web sources. The input consists of the query sentence and the intent, while the output is an expanded query.
\begin{tcolorbox}[colback=gray!10, boxrule=0pt]
\textbf{Module:} generator

\textbf{Task:} [Query Expansion] Expanding the original query according to the user's intent.

\textbf{Input:} [Query] what is the origin of Guiness beer [Intent] History

\textbf{Output:} [Expanded query] What is the origin of Guinness beer and what key historical events influenced its development?
\end{tcolorbox}

\subsubsection{Source Retrieval Instruction}
The objective of the source retrieval instruction is to invoke the web knowledge learned by the LLMs in order to generate the $r$ most closely related web page source names (where $1 \leq r \leq 5 $) based on the expanded query. The input is the expanded query, and the output consists of a specified number of website names.
\begin{tcolorbox}[colback=gray!10, boxrule=0pt]
\textbf{Module:} generator

\textbf{Task:} [source Retrieval] Retrieving five web sources that fulfill the user's requirement.

\textbf{Input:} [Query] What are some popular role-playing games for mobile devices?

\textbf{Output:} [Website] 1. IGN 2. Pocket Gamer 3. Android Authority 4. TouchArcade 5. AppAdvice
\end{tcolorbox}

\subsubsection{URL Identification Instruction}
The purpose of the URL recognition instruction is to convert the generated website names into accessible links. The input is the generated website names, and the output is the corresponding URLs for those website names.
\begin{tcolorbox}[colback=gray!10, boxrule=0pt]
\textbf{Module:} generator

\textbf{Task:} [URL Identification] Providing the homepage URL of the website. If the website does not exist, respond with 'This page does not exist.'

\textbf{Input:} [Website] National Technical Assistance Center on Transition

\textbf{Output:} [URL] https://transitionta.org/
\end{tcolorbox}

\subsubsection{Website Identification Instruction}
The objective of the URL recognition instruction is opposite to its purpose; the input consists of the retrieved URLs, and the output is the website names corresponding to those URLs.
\begin{tcolorbox}[colback=gray!10, boxrule=0pt]
\textbf{Module:} generator

\textbf{Task:} [Website Identification] Providing the website name of the homepage URL.

\textbf{Input:} [URL] https://www.ally.com/

\textbf{Output:} [Website] Ally Bank
\end{tcolorbox}

\subsection{Validation Instruction}
\subsubsection{Evidence Recognition Instruction}
The objective of this instruction is to locate evidence sentences from a lengthy text. The input consists of the query and the long text, and the output is the evidences.
\begin{tcolorbox}[colback=gray!10, boxrule=0pt]
\textbf{Module:} validator

\textbf{Task:} [Evidence Recognition] Identify evidence sentences from the document that can answer the query; if none exist, respond with 'No evidence.'

\textbf{Input:} [Query] at what age is medicare eligible [Document] The eligibility age has remained constant since 1965, when the program was established; however, only 38 percent of Americans said they would be eligible for Medicare at age 65, while one-third (32 percent) said they didn't know. 

\textbf{Output:} [Evidence] The eligibility age has remained constant since 1965.
\end{tcolorbox}

\subsubsection{Evidence Check Instruction}
The objective of evidence check instruction is to re-examine the confidence level of the evidence based on the foundation of evidence identification.

\begin{tcolorbox}[colback=gray!10, boxrule=0pt]
\textbf{Module:} validator

\textbf{Task:} [Evidence Check] Determine whether the evidence sentence can answer the query.

\textbf{Input:} [Query] types of von willebrand's disease [Evidence] Von Willebrand disease (VWD) is almost always inherited. Inherited means that the disorder is passed from parents to children though genes. You can inherit type 1 or type 2 VWD if only one of your parents passes the gene on to you. You usually inherit type 3 VWD only if both of your parents pass the gene on to you.

\textbf{Output:} [Result] Yes
\end{tcolorbox}

\subsubsection{Answer Generation Instruction}
The objective of this instruction is to produce answers with cited evidences. The input consists of a list of evidence, and the output is the answer.
\begin{tcolorbox}[colback=gray!10, boxrule=0pt]
\textbf{Module:} validator

\textbf{Task:} [Answer Generation] Generating answer that can answer the query based on references.

\textbf{Input:} [Query] How many electrons are able to fit in the third shell? Some answers say 8, others say 18 [References] [1] So the third shell can be considered to hold 8 or 18 electrons but in total the third shell can hold 18 electrons. [2] The third shell can carry up 18 electrons, but it is more stable by carrying only eight electrons.  [3] In my textbook, it says that the maximum number of electrons that can fit in any given shell is given by 2n². This would mean 2 electrons could fit in the first shell, 8 could fit in the second shell, 18 in the third shell, and 32 in the fourth shell.

\textbf{Output:} [Answer] The third shell can carry up to 18 electrons[1][2][3], but it is more stable by carrying only eight electrons[2].
\end{tcolorbox}

\subsection{Optimization Instruction}
\subsubsection{Source Correction Instruction}
The objective of this instruction update invalid sources. The input consists of the query, generated sources. The output is the updated web page sources.
\begin{tcolorbox}[colback=gray!10, boxrule=0pt]
\textbf{Module:} optimizer

\textbf{Task:} [Source Correction] Modifying retrieved sources based on user queries and suggestions.

\textbf{Input:} [Query] Top gaming laptops [Website] 1. PC Gamer 2. Tom's Hardware 3. IGN 4. Digital Trends 5. CNET [Suggestion] PC Gamer is inaccessible or does not include content that meets the needs of user

\textbf{Output:} [Corrected Website] 1. Tom's Hardware 2. IGN 3. Digital Trends 4. CNET 5. Laptop Mag
\end{tcolorbox}

\section{Case}

We display the output results of the model during the generation, verification, and optimization stages in Tables \ref{tab:case_generator}, \ref{tab:case_validator_check}, \ref{tab:case_validator_recognition}, and \ref{tab:case_optimizer}.


\begin{table*}[h]
  \caption{Examples of the model's source generation outcomes, we vary both $n$ and $r$ within the range of 1 to 3. \textnormal{Red indicates generated content that does not satisfy user requirements.}}
  \label{tab:case_generator}
  \begin{tabular}{p{4cm}|p{3cm}|p{6cm}|p{3cm}}
  \midrule
  Query & Intent & Expanded Query & Source \\
  \midrule
   Why did humans start shaving?  & (1) History of shaving & \textcolor[rgb]{0.8,0.2,0.1}{(1) What were the earliest methods of shaving and how did they evolve over time?} & (1) History.com\\
   \midrule
   Why are there only circular planets? Why can't there be square or rectangular ones? & (1) Scientific explanation for circular planets & (1) What is the scientific explanation for why planets are only circular and not square or rectangular? & (1) NASA \newline (2) Space.com\\
   \midrule
   The Lean Six Sigma Methodology & (1) Lean Six Sigma Methodology & (1) What are the key principles of the Lean Six Sigma Methodology and how can it be applied to different industries?  & (1) Lean Six Sigma \newline (2) Six Sigma \newline (3) \textcolor[rgb]{0.8,0.2,0.1}{Lean Enterprise Institute} \\
  \midrule
 Why does car sickness seem to hit the hardest when you look down at your phone, book, etc.? & (1) Causes of car sickness \newline(2) Effects of car sickness & (1) What are some common triggers of car sickness and how can you prevent it?\newline (2) What are the causes of car sickness and how can it be prevented? & (1) Healthline \newline (2) WebMD \\
 \midrule
 How do ads follow me from my phone to my computer? & (1) Ad tracking \newline(2) Ad targeting& (1) How do ads follow me from my phone to my computer?\newline (2) How do ads follow me from my phone to my computer and what are some ways to prevent them from tracking my online activity?& (1) TechRadar\newline (2) PCMag \newline(3) The New York Times\\
 \midrule
 Why are outlet store much cheaper than malls? & (1) Price comparison\newline (2) Shopping habits & (1) Why are outlet store prices much cheaper than mall prices?\newline (2) Why are outlet stores cheaper than malls and what are some other factors that influence the price of products in stores?&  (1) \textcolor[rgb]{0.8,0.2,0.1}{The Balance} \newline(2) Forbes \newline(3) Business Insider \newline(4) Consumer Reports\newline (5) \textcolor[rgb]{0.8,0.2,0.1}{NerdWallet}\\
 \midrule
 Why is it harder to fall asleep in noisy environments? & (1) Sleep deprivation\newline (2) Noise pollution\newline (3) Sleep disorders & (1) What are some ways to reduce noise levels in order to improve sleep quality? \newline(2) What are some ways to reduce noise pollution and improve sleep quality?\newline (3) What are some common sleep disorders and how can they be treated? & (1) Sleep Foundation\newline (2) National Sleep Foundation\\
 \midrule
 Why do companies have 'values' ? & (1) Purpose of company values\newline (2) Types of company values\newline (3) Examples of company values & (1) What are some examples of company values and how do they impact the workplace culture? \newline(2) What are some examples of company values and how do they impact business operations? \newline(3) What are some examples of company values and how do they impact the company's culture and operations? & (1) Harvard Business Review \newline(2) Forbes \\
 \midrule
 Explain: Integral Calculus & (1) Definition of integral calculus \newline(2) Examples of integral calculus\newline (3) Applications of integral calculus& (1) What is integral calculus and how is it used in mathematics and science? \newline(2) What are some examples of integral calculus in real-world applications? \newline(3) What are some real-world applications of integral calculus?& (1) Khan Academy \newline(2) Math Is Fun \newline (3) Math Open Reference\newline (4) Math.com \newline(5) Math.stackexchange\\

\bottomrule
  \end{tabular}
\end{table*}

\begin{table*}
  \caption{Examples of results returned using score-only validation strategy are presented. \textnormal{Red signifies the source is valid, but incorrectly identified invalid evidence.}}
  \label{tab:case_validator_check}
  \begin{tabular}{p{3cm}|p{2cm}|p{11cm}}
  \midrule
  Query & Source & Evidence \\ 
  \midrule
  Why doesn't it thunder during snow storms? & National Geographic& (1) The mystery behind thundersnow, a rar….The mystery behind thundersnow, a rare winter phenomenon. Until recent decades, we didn’t know if the phenomenon was even real.Now scientists are peeling back why thunder and lightning can happen in a snowstorm... \\
  \midrule
  \multirow{2}{3cm}{Why is a runners heart rate lower than a non runner}& \multirow{2}{2cm}{Runner's World} & (1) Runners and low heart rate.The “normal” heart rate range is usually between 60 bpm and 100 bpm and is most accurately measured when you first wake up before you begin to move around for the day... \\ 
  & &  (2) Generally getting fitter with aerobic exercise will lower the resting heart rate in most, depending on how much you are doing.We know that if you undertake regular aerobic exercise, probably more than three hours a week, then the heart will start to adapt and undergo physiological changes...\\
  \midrule
  \multirow{3}{3cm}{Why is ice-cold water so much more refreshing?} & \multirow{3}{2cm}{Healthline} & (1) Cold Water Therapy: Benefits of Cold Showers, Baths ... - Healthline.Cold immersion therapy can help reduce muscle soreness, cool you down after exercise, help your immune system and metabolism, and even improve your mood.But make sure your immersion in cold water is brief...\\
  & &  (2) Cold water therapy does, however, have some benefits that are proven by science.Let’s take a closer look at what these benefits are.Less muscle soreness Although researchers are debating the details, studies indicate that athletes who soak in cold water for short periods after exercise have less muscle soreness later on...\\
  & &   (3) \textcolor[rgb]{0.8,0.2,0.1}{If you have chronically dry skin, hot water can strip your sebum levels (oils) and exacerbate the issue, so cold water is a good alternative.}\\
  \midrule
  \multirow{2}{3cm}{PayPal, how it works, if it's trustworthy and should I get it?}& \multirow{2}{2cm}{PayPal}& (1) What are the benefits of using PayPal?.What are the benefits of using PayPal? PayPal is a fast, secure way to pay online.We help you make purchases at millions of online stores in the U.S. and across more than 200 global markets – all without the hassle of converting currency...\\
  & &  (2) \textcolor[rgb]{0.8,0.2,0.1}{The bank that approves the card (and lends them the cash to pay you) is called the issuing bank. payment systems offer convenient, quick, and secure transactions, leading to improved customer experiences, efficient record-keeping, and helpful integration capabilities...} \\
  \midrule
  \multirow{4}{3cm}{What are the point of Peanuts?} & \multirow{4}{3cm}{Healthline} & (1) Peanuts 101: Nutrition Facts and Health ….Peanuts provide protein, vitamins, minerals, and antioxidants.They may have other health benefits, including promoting fullness and helping protect against heart disease... \\
  & &  (2) Peanuts are low in carbs.This makes them a good dietary choice for people with diabetes.Peanuts are an excellent source of various vitamins and minerals.Peanuts are one of the richest dietary sources of biotin, which is important during pregnancy...\\
  & & (3) It helps your body’s cells convert carbs into energy and is essential for the function of your heart, muscles, and nervous system.- Phosphorus.Peanuts are a good source of phosphorus...\\
  & &(4) Peanuts and Diabetes: Benefits, Risks, and More.Eating peanuts may offer several benefits to people with type 2 diabetes, including aiding weight loss and lowering the risk of cardiovascular disease...\\
\bottomrule
  \end{tabular}
\end{table*}

\begin{table*}
  \caption{Examples of results returned using hybrid validation strategy are presented. \textnormal{Blue indicates the source is valid, but the model failed to find reliable evidence. Red signifies the source is valid, but incorrectly identified invalid evidence.}}
  \label{tab:case_validator_recognition}
  \begin{tabular}{p{3cm}|p{2cm}|p{11cm}}
  \midrule
  Query & Source & Evidence \\ 
  \midrule
  Why doesn't it thunder during snow storms? & National Geographic& \textcolor[rgb]{0.2,0.1,0.8}{No Evidence.} \\
  \midrule
  Why is a runners heart rate lower than a non runner& Runner's World & (1) The use of heart rate monitors is now common practice among many runners, but how much do we really know about resting heart rate? An indicator of aerobic fitness, a lower resting heart rate generally implies more efficient heart function. But are there exceptions and what warning signs should runners be looking out for?\\
  \midrule
  \multirow{2}{3cm}{Why is ice-cold water so much more refreshing?} & \multirow{2}{2cm}{Healthline} & (1) Cold immersion therapy can help reduce muscle soreness, cool you down after exercise, help your immune system and metabolism, and even improve your mood.\\
  & & \textcolor[rgb]{0.8,0.2,0.1}{(2) Cold water therapy does, however, have some benefits that are proven by science.} \\
  \midrule
 PayPal, how it works, if it's trustworthy and should I get it? & PayPal& (1) PayPal is a fast, secure way to pay online. We help you make purchases at millions of online stores in the U.S. and across more than 200 global markets all without the hassle of converting currency. It's free to sign up for an account and download...\\
 \midrule
  \multirow{2}{3cm}{What are the point of Peanuts?} &  \multirow{2}{3cm}{Healthline} & (1) According to an article in the British Journal of Nutrition , eating peanuts or peanut butter in the morning may help control your blood sugar throughout the day.\\
  & &  (2) Peanuts provide protein, vitamins, minerals, and antioxidants. They may have other health benefits, including promoting fullness and helping protect against heart disease.\\
\bottomrule
  \end{tabular}
\end{table*}


\begin{table*}
\begin{threeparttable}[b]
  \caption{Optimization results provided by different strategies. \textnormal{\textbf{Source} and \textbf{Updated Source} respectively denote the source lists outputted by the model's generator and optimizer. Red indicates the generated results are invalid.}}
  \label{tab:case_optimizer}
  \begin{tabular}{p{2cm}|p{2cm}|p{2cm}|p{3cm}|p{7cm}}
    \toprule
  Query &Source & Method & Updated source & Evidence \\
    \midrule
  \multirow{3}{2cm}{What is the "natural" sleeping position for humans} & \multirow{3}{2cm}{ \textcolor[rgb]{0.8,0.2,0.1}{(1) Sleep.org}} &Online & (1) Sleep Foundation & Sleeping on the right may increase pressure on your internal organs, which is why experts recommend the left for pregnant women and sleepers with acid reflux or gastroesophageal reflux disease (GERD).\\
  
\cline{3-5}

   &  &History Mining &  (1) Health Ambition & (1) Whether you’re a side, front or back sleeper, we all seem to naturally gravitate towards particular positions. Some people remain in one position; others change at different points \\
\cline{3-5}

   &  &Self-Critical & (1) Healthline &  (1) What's the Best Sleeping Position?It Depends.Different sleeping positions can make a difference on your sleep quality.You can keep a diary for a couple weeks to help determine what works\\
   \midrule
   \multirow{3}{2cm}{Why is eating burnt food bad for you?} & \multirow{4}{2cm}{\textcolor[rgb]{0.8,0.2,0.1}{(1) Mayo Clinic}}& Online & (1) Health Today &(1)  Before discussing the impact of eating burnt foods, it’s crucial to understand the process by which they’re created.Burnt foods are products of a chemical reaction known as the Maillard reaction.\\
   && &(2) Bon Appétit & (2) Plenty of research has come out suggesting she was right to be cautious of burnt foods, which produce various chemical compounds believed to cause cancer.\\
   \cline{3-5}
   & & History Mining & (1) Science Focus & (1) It’s long been known that just over-heating, let alone burning, some foods can lead to the formation of compounds linked to cancer.These include heterocyclic amines and so-called polycyclic aromatic hydrocarbons (PAHs), which can lead to fried or smoked foods...\\
   \cline{3-5}
   & & Self-Critical &(1) Healthline & (1) Burnt toast contains acrylamide, a compound formed in starchy foods during high-heat cooking methods like roasting, baking, and frying...\\
  \bottomrule
\end{tabular}
\end{threeparttable}
\end{table*}

\end{document}